\documentclass[3p,times,procedia]{elsarticle}
\usepackage{nupha_ecrc}

\volume{00}

\firstpage{1}

\journalname{Nuclear Physics A}

\runauth{V. Vovchenko et al.}

\jid{nupha}

\jnltitlelogo{Nuclear Physics A}

\usepackage{amssymb}

\usepackage[figuresright]{rotating}

\begin{document}

\begin{frontmatter}



\dochead{XXVIIth International Conference on Ultrarelativistic Nucleus-Nucleus Collisions\\ (Quark Matter 2018)}

\title{Lattice-based QCD equation of state at finite baryon density: Cluster Expansion Model}


\author[ITP,FIAS]{V.~Vovchenko}
\author[FIAS]{J.~Steinheimer}
\author[ITP]{O.~Philipsen}
\author[Wuppertal]{A.~P\'asztor}
\author[Wuppertal,Eotvos,Julich]{Z.~Fodor}
\author[Eotvos,ELTE]{S.D.~Katz}
\author[ITP,FIAS,GSI]{H.~Stoecker}

\address[ITP]{Institut f\"ur Theoretische Physik,
Goethe Universit\"at Frankfurt, D-60438 Frankfurt am Main, Germany}
\address[FIAS]{Frankfurt Institute for Advanced Studies, Giersch Science Center, D-60438 Frankfurt am Main, Germany}

\address[Wuppertal]{Department of Physics, Wuppertal University, D-42119 Wuppertal, Germany}

\address[Eotvos]{Institute for Theoretical Physics, E\"otv\"os University, H-1117 Budapest, Hungary}

\address[Julich]{J\"ulich Supercomputing Centre, Forschungszentrum J\"ulich, D-52425 J\"ulich, Germany}

\address[ELTE]{MTA-ELTE "Lend\"ulet" Lattice Gauge Theory Research Group, H-1117 Budapest, Hungary}
\address[GSI]{GSI Helmholtzzentrum f\"ur Schwerionenforschung GmbH, D-64291 Darmstadt, Germany}

\begin{abstract}
The QCD equation of state at finite baryon density is studied in the framework of a Cluster Expansion Model (CEM), which is based on the fugacity expansion of the net baryon density. 
The CEM uses the two leading  Fourier coefficients, obtained from lattice simulations at imaginary $\mu_B$, as the only model input and permits a closed analytic form.
Excellent description of the available lattice data at both $\mu_B = 0$ and at imaginary $\mu_B$ is obtained. 
We also demonstrate how the Fourier coefficients can be reconstructed from baryon number susceptibilities.
\end{abstract}


\end{frontmatter}


\section{Introduction}
\label{sec:intro}

Direct first-principle lattice QCD methods provide the equation of state of QCD only at zero chemical potential~\cite{Borsanyi:2013bia,Bazavov:2014pvz}, where a crossover-type transition is observed~\cite{Aoki:2006we}.
Lattice QCD calculations at finite $\mu_B$, on the other hand, are hindered by the sign problem.
Thermodynamic features of QCD at small but finite $\mu_B$ are therefore calculated using indirect methods, such as the reweighing techniques~\cite{Barbour:1997ej,Fodor:2001au}, the Taylor expansion around $\mu = 0$~\cite{Allton:2002zi}, the analytic continuation from imaginary $\mu$~\cite{deForcrand:2002hgr,DElia:2002tig}, or the canonical approach~\cite{Hasenfratz:1991ax,Nagata:2010xi}.

\begin{figure}[t]
  \centering 
  \includegraphics[width=.47\textwidth]{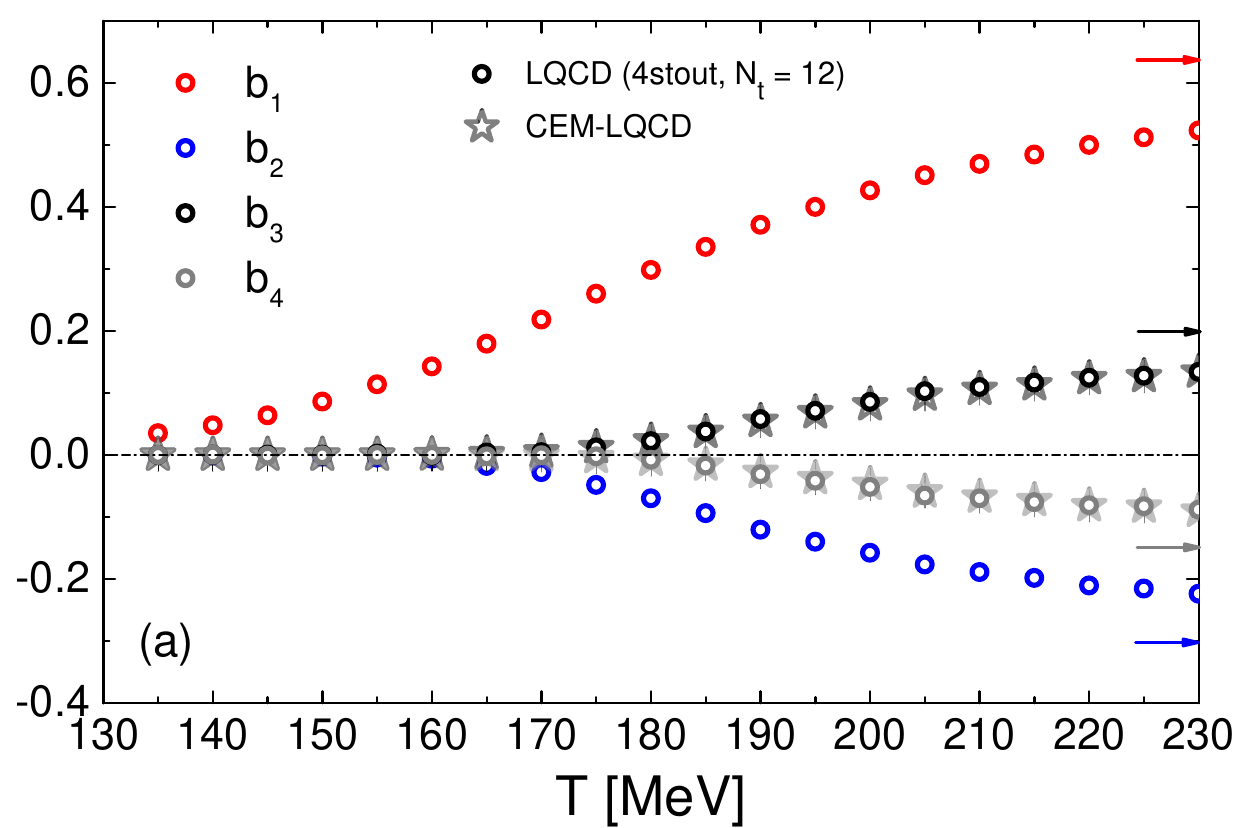}
  \includegraphics[width=.47\textwidth]{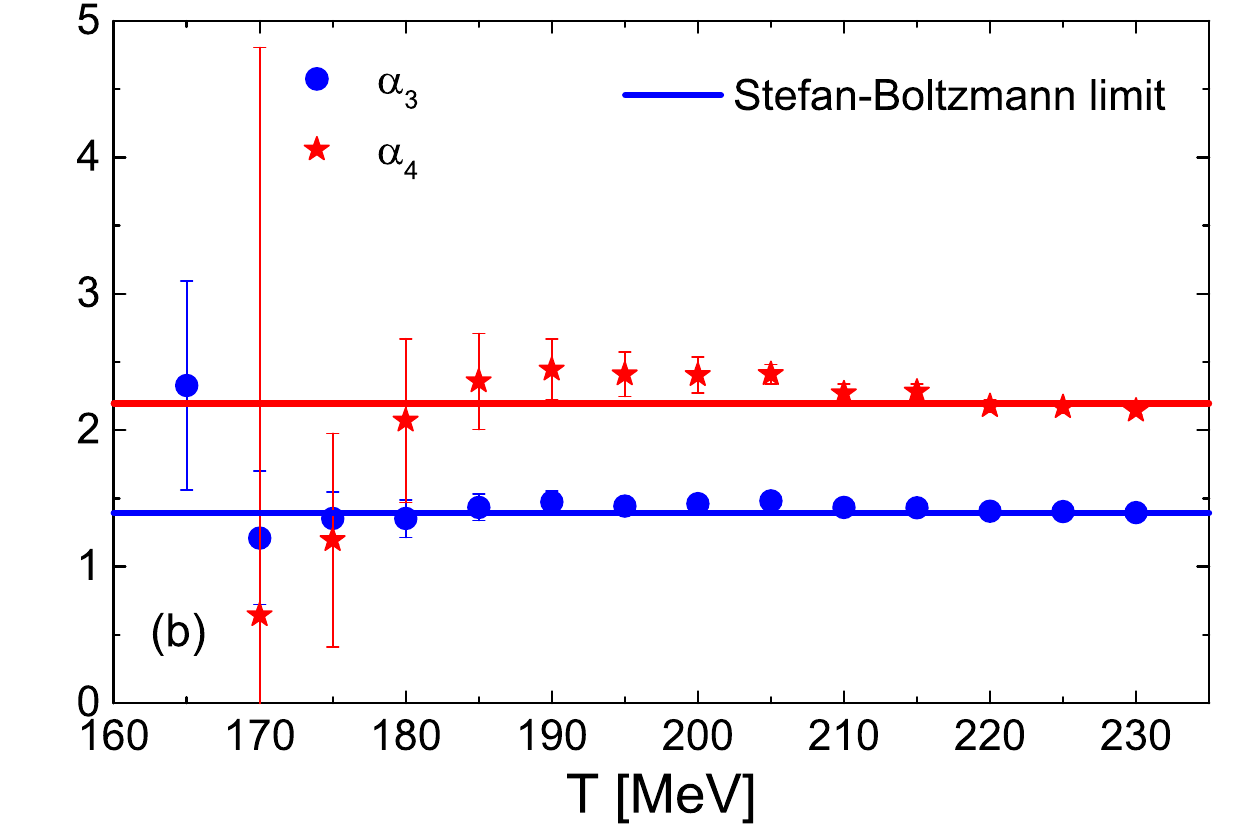}
  \caption{Temperature dependence of (a) the four leading Fourier coefficients from lattice QCD~\cite{Vovchenko:2017xad} and CEM-LQCD~\cite{Vovchenko:2017gkg}, and (b) the $\alpha_3$ and $\alpha_4$ ratios~[Eq.~(\ref{eq:alphak})], constructed from the lattice QCD data~\cite{Vovchenko:2017xad} and in the Stefan-Boltzmann limit of massless quarks~[Eq.~(\ref{eq:alphakSB})].
  } 
  \label{fig:alphas}
\end{figure}

The analysis presented here is based on the relativistic fugacity expansion of the net baryon density
\begin{equation}
\label{eq:fug}
\frac{\rho_B(T,\mu_B)}{T^3} = \sum_{k=1}^{\infty} b_k(T) \, \sinh(k\,\mu_B/T) \stackrel{\mu_B \to i \tilde{\mu}_B}{=} i \, \sum_{k=1}^{\infty} b_k(T) \, \sin(k\,\tilde{\mu}_B/T),
\end{equation}
which takes the form of a Fourier series at purely imaginary values of $\mu_B$. 
The leading four Fourier coefficients of the expansion~(\ref{eq:fug}) were recently calculated within imaginary $\mu_B$ lattice QCD simulations~\cite{Vovchenko:2017xad}, see Fig.~\ref{fig:alphas}a.

\section{Cluster Expansion Model}
\label{sec:cem}

\subsection{Higher-order coefficients from lower ones}

It was pointed out in Ref.~\cite{Vovchenko:2017xad} that the
  lattice behavior of the Fourier coefficients $b_2$, $b_3$, and $b_4$~(as well as baryon number susceptibilities~\cite{Vovchenko:2016rkn,Vovchenko:2017drx}) can be understood in terms of repulsive baryonic interactions.
The models for repulsive baryon-baryon interactions, one prominent example being the excluded volume model~\cite{Rischke:1991ke}, 
suggest the following ratios of Fourier coefficients to be temperature-independent:
\begin{equation}
\label{eq:alphak}
\alpha_k = \frac{[b_1(T)]^{k-2}}{[b_2(T)]^{k-1}} \, b_k(T), \qquad k = 3,4,\ldots
\end{equation}

The temperature dependence of the coefficients $\alpha_3$ and $\alpha_4$, constructed from the lattice QCD data~\cite{Vovchenko:2017xad}, is shown in Fig.~\ref{fig:alphas}b.
Both $\alpha_3$ and $\alpha_4$ show mild temperature dependence at $T>160$~MeV, whereas the uncertainties 
are too large to draw definitive conclusions for lower temperatures.
The available lattice data are consistent with temperature independent values for $\alpha_3$ and $\alpha_4$ calculated in the Stefan-Boltzmann limit of massless quarks:
\begin{equation}
\label{eq:alphakSB}
\alpha_k^{\rm SB} = 8^{k-1} \, \frac{(3+4\pi^2)^{k-2}}{(3+16\pi^2)^{k-1}} \, \frac{3+4\pi^2 k^2}{k^3}, \qquad k = 3,4,\ldots
\end{equation}

The above empirical observation forms the basis of the Cluster Expansion Model~(CEM)~\cite{Vovchenko:2017gkg}. 
The CEM takes $b_1(T)$ and $b_2(T)$ as an input and assumes that all higher-order Fourier coefficients are given by
\begin{equation}
\label{eq:bkCEM}
b_k (T) = \alpha_k^{\rm SB} \, \frac{[b_2(T)]^{k-1}}{[b_1(T)]^{k-2}}, \qquad k = 3,4,\ldots
\end{equation}
This relation neglects the \emph{connected} 3-baryon 
correlations and thus is valid for a sufficiently dilute system.

\subsection{Analytic form}

The fugacity expansion in Eq.~(\ref{eq:fug}) can be analytically summed for the CEM ansatz~(\ref{eq:bkCEM}).
The result is
\begin{equation}
\label{eq:analyt}
\frac{\rho_B(T,\mu_B)}{T^3} = 
-\frac{2}{27 \pi^2} \, \frac{\hat{b}_1^2}{\hat{b}_2} \, \left\{ 4 \pi^2  \, \left[ \textrm{Li}_1\left(x_+\right) - \textrm{Li}_1\left(x_-\right) \right] + 3 \, \left[ \textrm{Li}_3\left(x_+ \right) - \textrm{Li}_3\left(x_-\right) \right] \right\}.
\end{equation}
Here $\hat{b}_{1,2} = \displaystyle \frac{b_{1,2}(T)}{b_{1,2}^{\rm SB}}$, $x_{\pm} = \displaystyle - \frac{\hat{b}_2}{\hat{b}_1} \, e^{\pm \mu_B/T}$, and $\textrm{Li}_s(z) = \displaystyle \sum_{k=1}^{\infty} \frac{z^k}{k^s}$ is the polylogarithm.

It follows from 
Eq.~(\ref{eq:analyt}) that there is no singular behavior in the CEM at real values of the baryochemical potential $\mu_B$, provided that $b_1 > 0$ and $b_2 < 0$ as suggested by the lattice data for $T > 135$~MeV~\cite{Vovchenko:2017xad}.
The CEM corresponds to a no-critical-point scenario for the QCD phase diagram.
\emph{Therefore, any unambiguous signal of the QCD critical point will show up as a deviation from the CEM result.}
However, the CEM does contain singularities in the complex $\mu_B / T$ plane, due to the polylogarithm, and this has certain consequences for the radius of convergence of  Taylor expansion around $\mu_B / T = 0$~(see Ref.~\cite{Vovchenko:2017gkg} for details).

\section{Results}
\label{sec:results}

\subsection{Baryon number susceptibilities}

The baryon number susceptibilities $\chi_k^B = \partial^{k-1}(\rho_B/T^3) / \partial(\mu_B/T)^{k-1}$ in the CEM read
\begin{equation}
\label{eq:chikCEM}
\chi_k^B (T, \mu_B) = -\frac{2}{27 \pi^2} \, \frac{\hat{b}_1^2}{\hat{b}_2} \, \left\{ 4 \pi^2  \, \left[ \textrm{Li}_{2-k}\left(x_+\right) + (-1)^k \, \textrm{Li}_{2-k}\left(x_-\right) \right] + 3 \, \left[ \textrm{Li}_{4-k}\left(x_+ \right) + (-1)^k \, \textrm{Li}_{4-k}\left(x_-\right) \right] \right\}.
\end{equation}

Leading order baryon number susceptibilities at $\mu_B = 0$ have recently been computed in lattice QCD~\cite{Bellwied:2015lba,DElia:2016jqh,Bazavov:2017dus,Bazavov:2017tot,Borsanyi:2018grb}.
A comparison with these lattice data can test the predictive power of the CEM.

\begin{figure}[t]
  \centering 
  \includegraphics[width=\textwidth]{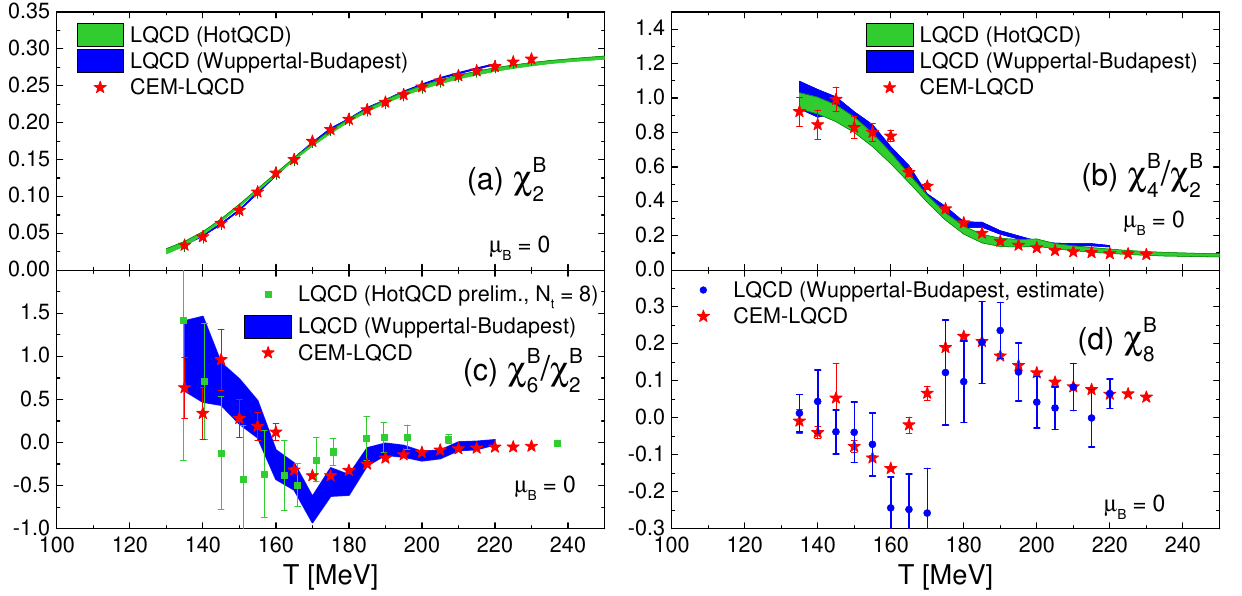}
  \caption{Temperature dependence of the net baryon susceptibilities (a) $\chi_2^B$, (b) $\chi_4^B / \chi_2^B$, (c) $\chi_6^B / \chi_2^B$, and (d) $\chi_8^B$, calculated within CEM-LQCD (red stars). Lattice QCD data of Wuppertal-Budapest~\cite{Borsanyi:2018grb} and HotQCD~\cite{Bazavov:2017dus,Bazavov:2017tot} collaborations are shown by the blue and green bands/symbols, respectively.
  } 
  \label{fig:chik}
\end{figure}

Figure \ref{fig:chik} depicts the temperature dependence of $\chi_2^B$, $\chi_4^B/\chi_2^B$, $\chi_6^B / \chi_2^B$, and $\chi_8^B$, calculated in CEM  and compared to the lattice data of Wuppertal-Budapest~\cite{Borsanyi:2018grb} and HotQCD collaborations~\cite{Bazavov:2017dus,Bazavov:2017tot}.
The CEM calculations use the Wuppertal-Budapest data~\cite{Vovchenko:2017xad} for $b_1(T)$ and $b_2(T)$ as an input and are therefore labeled CEM-LQCD in Fig.~\ref{fig:chik}.
CEM results are in quantitative agreement with the lattice data for $\chi_2^B$ and $\chi_4^B / \chi_2^B$.
The CEM is also consistent with the lattice data for $\chi_6^B/\chi_2^B$ and $\chi_8^B$, although these data are still preliminary and have large error bars.
One interesting qualitative feature is the dip in the temperature dependence of $\chi_6^B / \chi_2^B$, where this quantity is negative.
It was interpreted as a possible signature of chiral criticality~\cite{Friman:2011pf}.
Given that this behavior is also present in CEM~(see red stars in Fig.~\ref{fig:chik}c), i.e. in a model which has no critical point, we conclude that the negative dip in $\chi_6^B / \chi_2^B$ cannot be considered as an unambiguous signal of chiral criticality.

\subsection{Reconstructing the Fourier coefficients $b_1$ and $b_2$ from susceptibilities}

\begin{figure}[t]
  \centering 
  \includegraphics[width=.49\textwidth]{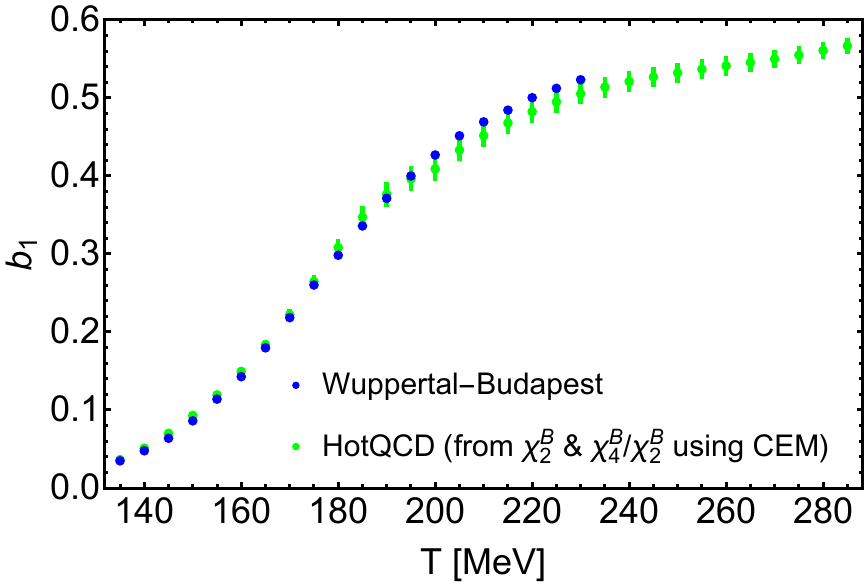}
  \includegraphics[width=.49\textwidth]{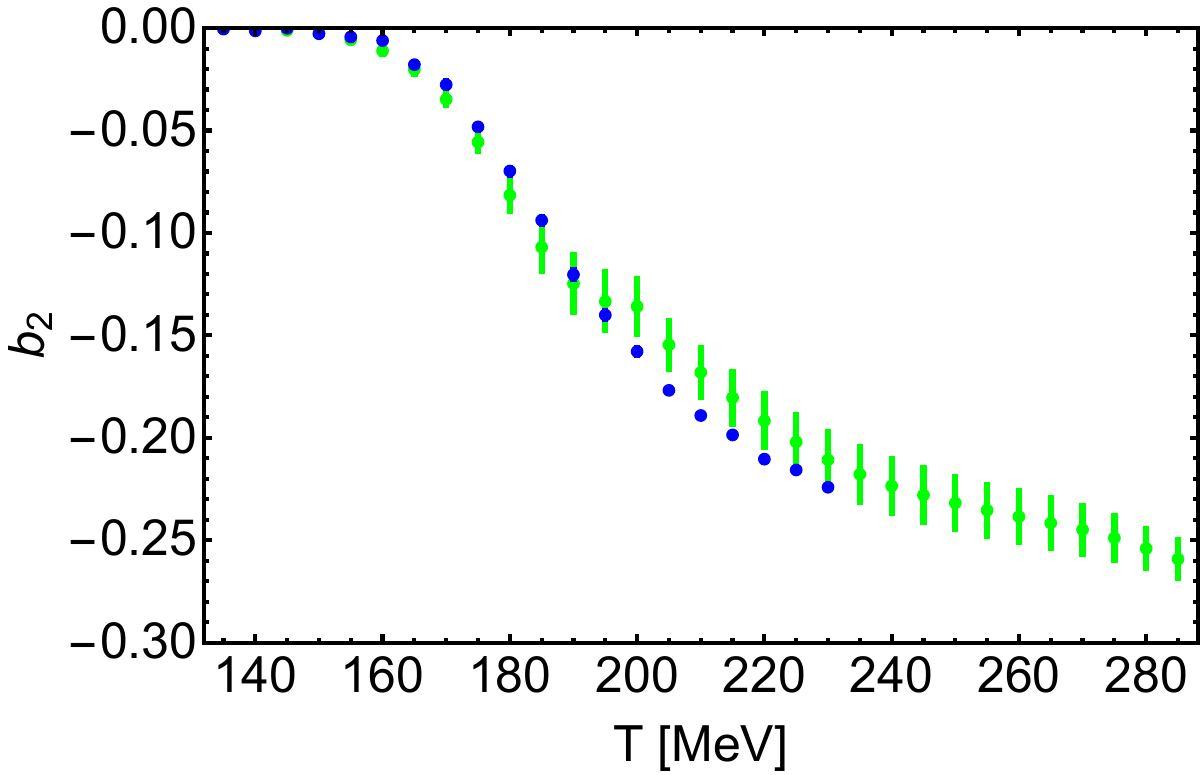}
  \caption{Temperature dependence of the leading two Fourier coefficients $b_1(T)$ and $b_2(T)$, calculated in lattice QCD simulations by the Wuppertal-Budapest collaboration~\cite{Vovchenko:2017xad}, and reconstructed from the lattice data of the HotQCD collaboration~\cite{Bazavov:2017dus,Bazavov:2017tot} for $\chi_2^B$ and $\chi_4^B/\chi_2^B$ using CEM~[Eq.~(\ref{eq:chikCEM})].
  } 
  \label{fig:b1b2HotQCD}
\end{figure}

All baryon number susceptibilities at a given temperature are determined in the CEM by two parameters -- the leading two Fourier coefficients $b_1$ and $b_2$.
One can now consider a reverse prescription -- assuming the validity of the CEM ansatz one can extract the values of $b_1$ and $b_2$ at a given temperature from two independent combinations of baryon number susceptibilities by reversing Eq.~(\ref{eq:chikCEM}).
We demonstrate this by considering the lattice QCD data of the HotQCD collaboration for $\chi_2^B$ and $\chi_4^B / \chi_2^B$.
The temperature dependence of the $b_1$ and $b_2$ coefficients, reconstructed from the HotQCD collaboration's lattice data on the basis of CEM~[Eq.~(\ref{eq:chikCEM})], is shown in Fig.~\ref{fig:b1b2HotQCD} by the green symbols.
The extracted values agree rather well with the imaginary $\mu_B$ data of the Wuppertal-Budapest collaboration, shown in Fig.~\ref{fig:b1b2HotQCD} by the blue symbols.
This agreement can be regarded as a possible implicit evidence for both, the consistency between the lattice results of the Wuppertal-Budapest and HotQCD collaborations, and that the CEM ansatz provides an accurate description for all observables considered here.

\section{Summary}

We presented the Cluster Expansion Model for the QCD equation of state at finite baryon density, which is based on the relation~(\ref{eq:bkCEM}) between higher-order and the leading two Fourier coefficients of the net baryon density, suggested by the recent lattice data at imaginary $\mu_B$.
The analytic structure of the CEM has no critical point, therefore unambiguous signals of the hypothetical QCD critical point in various observables must show up as deviations from CEM predictions.
The presently available lattice data on Fourier coefficients and baryon number susceptibilities do not show such deviations.
Given its simplicity and consistency with the lattice data, the CEM based equation of state can be useful for hydrodynamic simulations of heavy-ion collisions at finite baryon density.





\bibliographystyle{elsarticle-num-mod}
\bibliography{vovchenko_qm18}







\end{document}